\documentclass[twocolumn,amsmath,amssymb,aps,pra,superscriptaddress]{revtex4-1}
\usepackage{amsmath,amssymb}
\usepackage{microtype}
\usepackage{graphicx}
\usepackage{physics}
\usepackage{subfigure}
\usepackage{mathrsfs}
\usepackage[normalem]{ulem}

\usepackage{hyperref}
\newcommand{\beq}{\begin{equation}}
\newcommand{\eeq}{\end{equation}}
\newcommand{\beqa}{\begin{eqnarray}}
\newcommand{\eeqa}{\end{eqnarray}}
\usepackage{lipsum}

\usepackage{bm}
\usepackage{color}
\usepackage{xcolor}
\usepackage[utf8]{inputenc}
\usepackage{hyperref}
\usepackage{mathtools}
\usepackage{natbib}
\usepackage{subfigure}
\usepackage{url}
\usepackage{graphicx}
\usepackage{subfigure}
\usepackage{enumitem}
\usepackage{epstopdf}
\usepackage{xcolor}
\usepackage{ulem}

\hypersetup{
    colorlinks=true,
    linkcolor=blue,
    filecolor=magenta,
    urlcolor=purple,
    citecolor=purple
}

\usepackage{xcolor}
\usepackage{soul}

\begin{document}
\title{Universal properties of dipolar Bose polarons in two dimensions}

\author{J. S\'anchez-Baena}
  \affiliation{Departament de
  F\'{\i}sica, Campus Nord B4-B5, Universitat Polit\`ecnica de
  Catalunya, E-08034 Barcelona, Spain}
\author{L. A. Pe\~na Ardila}
  \affiliation{School of Science and Technology, Physics Divisio, University of Camerino,
  Via Madonna delle Carceri, 9B - 62032 (MC), Italy}
\author{G. E. Astrakharchik}
  \affiliation{Departament de
  F\'{\i}sica, Campus Nord B4-B5, Universitat Polit\`ecnica de
  Catalunya, E-08034 Barcelona, Spain}
\author{F. Mazzanti}
  \affiliation{Departament de
  F\'{\i}sica, Campus Nord B4-B5, Universitat Polit\`ecnica de
  Catalunya, E-08034 Barcelona, Spain}

\begin{abstract} 
  We study the quasiparticle properties of a dipolar impurity immersed
  in a two-dimensional dipolar bath. We use the ab-initio Diffusion
  Monte Carlo technique to determine the polaron energy, effective
  mass and quasiparticle residue.  We find that both the polaron energy and quasiparticle residue
  follow a universal behaviour with respect to the polarization angle when properly scaled in terms
  of the scattering length.
  This trend is maintained over a wide range of values of the gas parameter,
  even in the highly correlated regime. Instead, the effective mass shows growing anisotropy as the tilting angle is increased, which is induced, mainly, by the anisotropy of the impurity-boson dipole-dipole interaction. Surprisingly, the effective mass is larger in the direction of minimum inter-particle repulsion.
  Finally, we use our Monte Carlo results to
  check the accuracy of perturbative approaches and determine their
  range of validity in terms of the gas parameter.
\end{abstract}

\date{\today}

\maketitle


\section{Introduction}

Impurities interacting with a complex quantum-many-body environment have been the subject of intense research in recent years. In the solid-state realm, impurities interacting with an ionic crystal disrupt the media and are screened by  lattice phonons, forming quasiparticles known as {\em polarons}~\cite{Landau48}. Polarons have been found to play an important role in semiconductor transport~\cite{PolaronsandExcitonsBook}, colossal magnetoresistance~\cite{MagnetoresistanceBook}, as well as non-equilibrium phenomena such as quantum heat transport~\cite{Hsieh19}. The high tunability and controllability of ultracold quantum gases offers an excellent platform to probe polaron physics in a clean environment, which has motivated a considerable amount of experimental~\cite{Kohstall:2012kg,nils16,corson16,ardila19,zoe20,etrych24,skou2021,cayla23,oppong19} and theoretical~\cite{braaten10,liu20,liu20b,yegovtsev22,massignan21,yoshida18,yegovtsev23,ArdilaPRR2020,yegovtsev23b,fujii22,nakano24,tiene24,volosniev23,nishimura21,guebli19,bombin19,bombin21,wenzel18,Ardila:2018jo,BenKain:2014he,matveeva13,camacho23,yegovtsev23b} works. Polarons have been experimentally realized in bosonic~\cite{nils16,corson16,ardila19,zoe20,etrych24,skou2021,cayla23} as well as fermionic~\cite{oppong19,Kohstall:2012kg} environments, and state-of-the art techniques like rf spectroscopy or the measurement of driven Rabi oscillations allow to probe quasiparticle properties like the polaron energy and the quasiparticle residue, respectively.

On the other hand, dipolar systems present rich physics due to the unique combination of traits of the dipole-dipole interaction (DDI): its anisotropy and long range character in three dimensions. This mix gives rise to a wide variety of phenomena, such as the emergence of ultra-dilute self-bound droplets~\cite{Pfau:nature:2016,Pfau:nature2:2016,Pfau:PRL:2016,ferlaino16,Mazzanti:PRL:2016,bottcher19}, supersolids~\cite{Pfau:PRX:2019,Ferlaino:PRX:2019,Tanzi:Nature:2019,Tanzi:Science:2021,Biagioni:PRX:2022,Modugno:PRL:2019,Bland2022,zhang19,zhang21,hertkorn21,maucher24,baena23shell,macia12,macia14,gallemi22,ferlaino20,smith22} which may be even self-bound in the case of a dipolar mixture~\cite{arazo23}, striped liquids~\cite{mazzanti23} and the anomalous emergence of supersolidity upon increasing temperature~\cite{Ferlaino:PRL:2021,baena22,baena24}, among others. In the context of the polaron problem, an immediate question emerges: how do the unique properties of the DDI affect the quasiparticle properties and dynamics of an impurity immersed in a dipolar medium? Several theoretical works have addressed this question in a variety of different conditions: from non-dipolar~\cite{nakano24,nishimura21} and dipolar~\cite{bombin19} impurities immersed in a dipolar fermionic medium, to an impurity-medium bilayer configuration~\cite{matveeva13,tiene24} or dipolar~\cite{volosniev23,wenzel18,Ardila:2018jo} as well as non-dipolar~\cite{BenKain:2014he,guebli19} impurities immersed in a dipolar BEC. It has also been shown that dipolar impurities can potentially function as tools to probe the properties of dipolar bosonic quantum droplets due to their neglectable back-action on the droplets~\cite{wenzel18}. In almost all cases, though, theoretical studies are restricted to an ultra-dilute bath, meaning that the characterization of the polaron properties for large gas parameters of the background is still a rather unexplored subject.

Precisely when the density (and thus, the correlations) of the medium are increased, and the inter-particle distances become of the order of the range of interactions, it becomes relevant to determine the regime for which a universal description of the problem is quantitatively accurate. In ultra-dilute conditions where the impurity-medium interaction is short range, both the ground state properties and the dynamics of the impurity are expected to be universal~\cite{braaten10,yoshida18,skou2021,fujii22,etrych24}, that is, dependent only on the scattering length $a$ and the density $n$. However, for impurity-bath interactions in the unitary limit~\cite{massignan21,yegovtsev22} or for a sufficiently dense background~\cite{bombin19,bombin21}, universality is expected to be lost at some point, where the description of the system becomes dependent on the details (range) of the interactions~\cite{bombin21}. For instance, for the case of an impurity immersed on a Fermi gas in two dimensions, the universality in the polaron energy is lost for gas parameters as low as $x=10^{-5}$, while the quasiparticle residue remains universal up to $x=10^{-2}$~\cite{bombin19}.

In this sense, the two-dimensional dipolar system in the absence of an impurity shows a universal behaviour of its ground state properties with respect to the polarization angle of the dipoles for surprisingly large values of the gas parameter~\cite{SanchezBaena2022}. That is, when dipoles are tilted with respect to the perpendicular direction of the plane and the DDI changes, the scaled energy $E m a^2/ N \hbar^2$, scaled pair-distribution function $g(\vec{r}/a)$ and condensate fraction depend only on the gas parameter $x=n a^2$. This is the case even in the ultra-correlated regime ($x \sim 100$). In this sense, this is a similar universality to the one presented in Ref.~\cite{durst24} for a Rydberg impurity, where its absorption spectra for several Rydberg interactions (for different values of the principal quantum number) is the same as long as $x=n a^3$ is kept constant.

\begin{figure}[h]
\includegraphics[width=\linewidth]{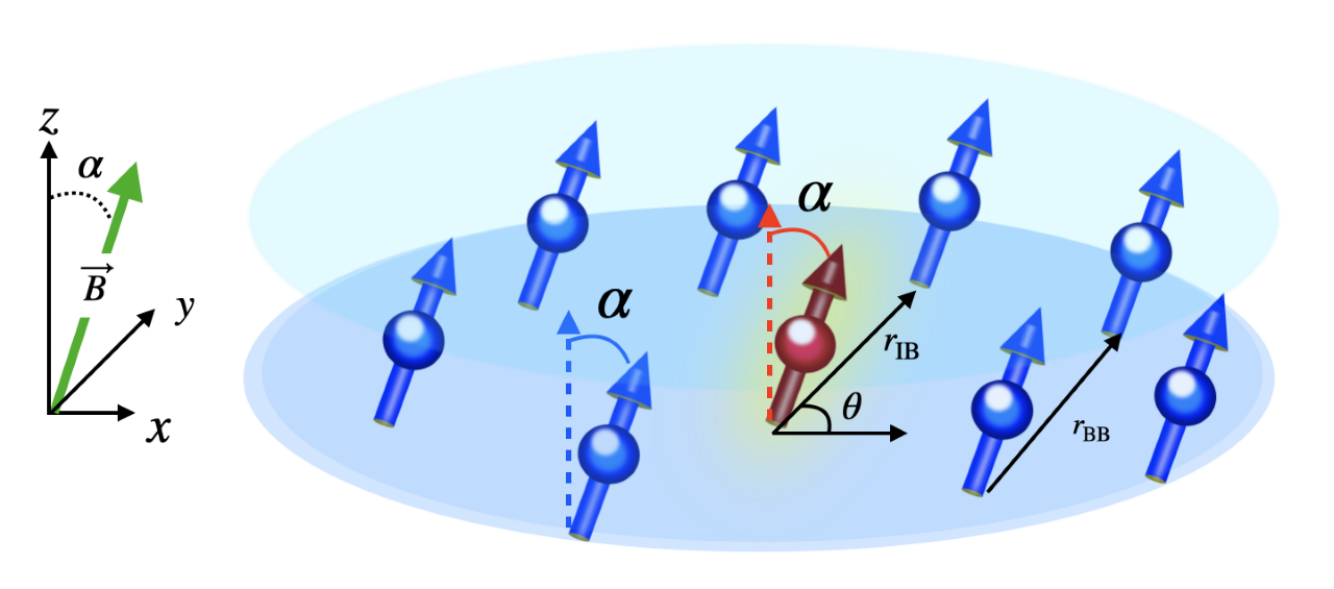}
\caption{Sketch of the system.
Blue arrows depict dipolar atoms from the bath confined to the $x-y$ plane and the red arrow shows the impurity.
An external magnetic field is used to polarize all atoms and the impurity in the same direction, forming the tilting angle $\alpha$ with the respect to the direction normal to the plane.
The azimuthal angle $\theta$ encodes the anisotropy of the system.
}
\label{fig:Cover}
\end{figure}

In light of this unexpected universal behaviour displayed by the bulk system, we address the question of whether this anomalous universality translates to the polaron properties once an impurity is introduced into the system. For that, we compute the quasiparticle properties (polaron energy, quasiparticle residue and effective mass) of the repulsive Bose polaron as a function of the polarization angle of the dipoles for different values of the gas parameter of the bulk, reaching up to the very strongly correlated regime. We do so by means of the Diffusion Monte Carlo (DMC) method. We set all particles (the
impurity and the ones in the bath) polarized along the same
direction in space.
The dipolar interaction depends on the angle $\alpha$
formed by the polarization field and the $z$-axis, and the dipolar strength $C_{dd}$. Denoting by I and B the impurity and a background atom,
respectively, the dipolar interaction reads
\begin{eqnarray}
V_{\sigma\sigma'}(\mathbf{r}_{i j}) = 
C_{dd}^{\sigma\sigma'}
\left(
{1-3\sin^{2}\alpha\cos^{2}\theta_{ij}
\over
r_{ij}^3}
\right) \ ,
\label{eq:V12}
\end{eqnarray}
where $\sigma, \sigma'\in\{I,B\}$.
In this expression  $\mathbf{r}_{i j}=\mathbf{r}_i-\mathbf{r}_j$
is the in-plane relative position vector between any pair of
atoms, while $r_ {ij}=|{\bf r}_i - {\bf r}_j|$
and $\theta_{ij}$ are the corresponding distance and relative
orientation angle, respectively, as shown in Fig.~\ref{fig:Cover}.
Furthermore, $C_{dd}^{\sigma\sigma'}=6\pi\hbar^{2}\sqrt{d_
  {\sigma}d_{\sigma'}}/\mu^{\sigma\sigma'}$, with $d_{\sigma} = m_{\sigma}
C_{dd}^{\sigma\sigma}/ (12 \pi \hbar^2)$ the corresponding dipolar
length, and $\mu^{\sigma\sigma'} = m_{\mathrm{\sigma}}m_{\mathrm{\sigma'}}/(m_{\mathrm{\sigma}}+m_
{\mathrm{\sigma'}})$ the reduced mass between two atoms. The expression in Eq.~(\ref{eq:V12}) shows that the dipolar
interaction is anisotropic and depends on both the polarisation angle
$\alpha$ and the interaction strength $C_{dd}$, both of which
determine the scattering length. The opposite is also true, namely,
that a given a set of particles feel a different dipolar interaction
when either $\alpha$ or the scattering length is changed.

\section{System and numerical method}

The system consists of a single impurity interacting with a background
of bosonic dipoles in two dimensions at fixed density $n$ and at zero
temperature, described by the Hamiltonian
\begin{equation}
  \mathcal{H}=-\frac{\hbar^{2}}{2m_{B}}\sum_{i=1}^{N}\nabla_{i}^{2} -
  \frac{\hbar^{2}}{2m_{I}}\nabla_{I}^{2} +
  \sum_{i<j}V_{\mathrm{BB}}\left({\bf r}_{ij}\right) +
  \sum_{i=1}^{N}V_{\mathrm{IB}}\left({\bf r}_{iI}\right) \ .
\label{eq:HFull}
\end{equation}
The first two terms in this expression represent the kinetic energy of
the host bath and the impurity, 
while the last ones correspond to the dipolar interactions between
the background atoms and with the impurity, respectively.

In the following we consider the impurity and background bosons to
have the same mass, so we set 
$m_{\mathrm {I}}=m_{\mathrm{B}}=m$. This assumption is well suited,
for example, when we consider the atoms in the bath and the impurity
to be different isotopes of the same highly dipolar, heavy atom  as could be $^{162}$Dy and $^{164}$Dy.  This is also a
realistic assumption when the impurity and the background particles
correspond to the same isotopes, but in different hyperfine states.
We also restrict the analysis to tilting angles $\alpha \in [0, 0.615]$ rad, as for larger values one has that $1<3\sin^{2}\alpha$, and thus the DDI ceases to be repulsive for all $\mathbf{r}_{i j}$, which induces a collapse into the system in the absence of
additional hard core repulsive forces.
%
%

Within this model, the $s$-wave scattering length
for both the \{I,B\} and \{B,B\} interaction pairs 
become~\cite{macia11}
\begin{equation}
  a_{\sigma\sigma'}(\alpha,C_{dd}^{\sigma\sigma'}) \simeq
  \frac{mC_{dd}^{\sigma\sigma'}}{4\pi\hbar^{2}}\exp(2\gamma)
  \left(1-\frac{3\sin^{2}\alpha}{2}\right)
  \label{scatt_length}
\end{equation}
where $\gamma$ is the Euler-Mascheroni constant $\gamma =
0.577\cdots$. These values fix another relevant parameter of the system,
$\beta=C_{dd}^{\mathrm{IB}}/C_{dd}^{\mathrm{BB}} =
a_{\mathrm{IB}}/a_{\mathrm{BB}}$, which sets the relative strength
between the impurity-background and the background-background
interactions.
In this way, the system properties are governed by $\alpha$, $\beta$
and the density $n$.

We perform the calculations using the Diffusion Monte Carlo method~\cite{chin90}. In DMC, one numerically implements the imaginary time evolution equation
\begin{align}
 &\psi_T(\mathbf{R}) \psi(\mathbf{R}, \tau + \Delta \tau) = \nonumber \\
 &\int d\mathbf{R'} G(\mathbf{R},\mathbf{R'},\Delta \tau) \frac{\psi_T(\mathbf{R})}{\psi_T(\mathbf{R'})} \psi_T(\mathbf{R'}) \psi(\mathbf{R'}, \tau) \ , \label{imag_ev}
\end{align}
where $\tau = i t/\hbar$ is the imaginary time, $G(\mathbf{R},\mathbf{R'},\Delta \tau) = \bra{\mathbf{R}} \exp \left( - \hat{H} \Delta \tau \right) \ket{\mathbf{R'}}$ is the imaginary time Green's function and $\psi_T(\mathbf{R})$ is the trial wave function, an input to the method. This trial wave function helps reduce the variance of the estimations if chosen appropriately. To numerically implement Eq.~\ref{imag_ev}, one first obtains a statistical representation of the density associated to a given trial wave function, $\rho_T(\mathbf{R}) = \abs{\psi_T(\mathbf{R})}^2$. This representation consists of a set of points in coordinate space named \textit{walkers} (i.e. $\{ \vec{R} \}_i = \{ \vec{r}_1\text{, ..., } \vec{r}_N\}_i$ where $i$ is the walker index). A set of transformations are then applied to the walkers, such that, at the end of the process, they statistically represent the probability distribution $\psi_0(\mathbf{R}) \psi_T(\mathbf{R})$, with $\psi_0(\mathbf{R})$ the ground state wave function of the system. These transformations are obtained by interpreting the quantity $G(\mathbf{R},\mathbf{R'},\Delta \tau) \frac{\psi_T(\mathbf{R})}{\psi_T(\mathbf{R'})}$ as a probability distribution. Observables are estimated as
\begin{align}
 \langle \hat{O} \rangle = \frac{\int d\mathbf{R} \text{ } \psi_T(\mathbf{R}) \hat{O} \psi_0(\mathbf{R}) }{ \int d\mathbf{R} \text{ } \psi_T(\mathbf{R}) \psi_0(\mathbf{R})} \ .
\end{align}
Notice that any quantity that commutes with the Hamiltonian, such as the energy, can be computed exactly (up to statistical uncertainty) regardless of the choice of the trial wave function. We use a trial wave
function of the Jastrow form
\begin{equation}
\psi_T(\mathbf{R})=\prod_{i=1}^N f_{\rm IB} \left(\mathbf{r}_{i I}\right)
\prod_{i<j} f_{{\rm BB}}\left(\mathbf{r}_{i j}\right) \ ,
\label{trial_Psi}
\end{equation}
where the two-body correlation factors $f_\mathrm{IB}$
(impurity-background) and $f_\mathrm{BB}$ (background-background) are
obtained from the solution of the zero-energy two-body problem, as
done in previous works~\cite{SanchezBaena2022, macia11}.  These
functions have been matched with suitable large-distance phononic
tails in order to recover the proper behavior of the many-body wave
function.




\begin{figure}
\includegraphics[width=0.45\textwidth]{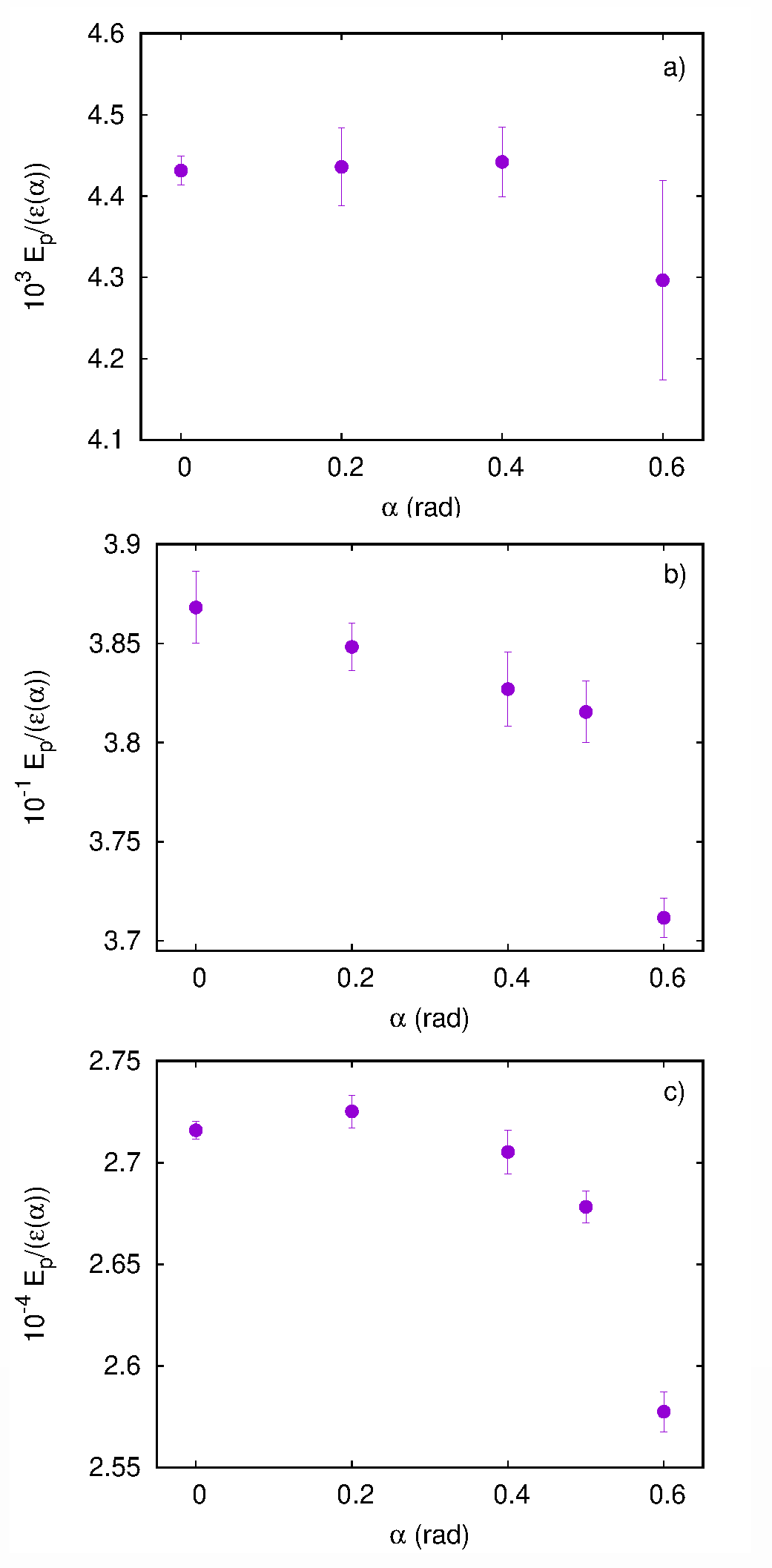}
\caption{ DMC results for the polaron energy in scattering length units (i.e $\epsilon(\alpha) = \hbar^2/(m a_{\rm BB}^2(\alpha))$, with $a_{\rm BB}(\alpha))$ defined in Eq.~\ref{scatt_length}) for $x=0.001$ (a), $x=1$ (b) and $x=100$ (c). }
\label{fig:fig2}
\end{figure}

\begin{figure}
\includegraphics[width=0.45\textwidth]{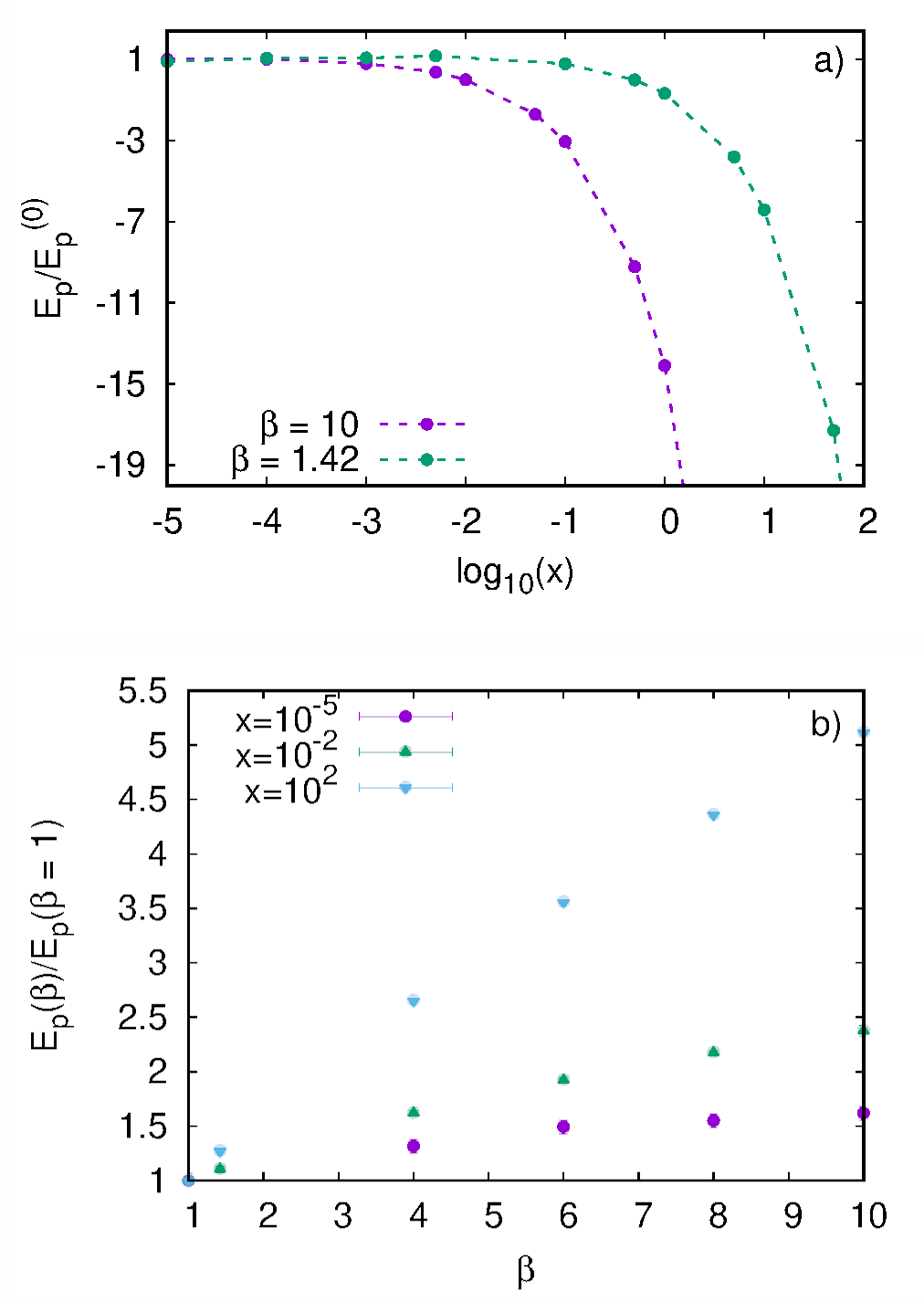}
\caption{(a) Ratio
  between the DMC ($E_p$) and the first order perturbation theory
  ($E_p^{(0)}$, see Eq.~\ref{Emf_polaron}) polaron energies for $\beta=1.42$ and $\beta=10$. (b)
  Polaron energy as a function of the impurity-boson coupling $\beta=C_{dd}^{\mathrm{IB}}/C_{dd}^{\mathrm{BB}}$ (dimensionless)
  for different values of the gas parameter. Energies have been rescaled with
  respect to its corresponding value at $\beta=1$.}
\label{fig:fig3}
\end{figure}

\section{Results}

In the following, we study how the polaronic properties (polaron energy, quasiparticle residue and effective mass) and the pair-correlation function depend on the density $n$ and the tilting angle of the dipoles $\alpha$ for a fixed ratio $\beta$. The dependence on $\beta$ is reported for the polaron energy.

\subsection{Polaron energy and pair-distribution function}

The driving quantity in any DMC calculation is the ground state
energy, which therefore becomes in a natural way the first property to
analyze. For a dilute system, the mean field prediction for the
polaron energy~\cite{ArdilaPRR2020}
\begin{equation}
  E_p^{(0)} =
  -\frac{4 \pi n \hbar^2}{m \ln(n a_{\mathrm{IB}}^2)} \ ,
  \label{Emf_polaron}
\end{equation}
is expected to hold, irrespective of the details of the
interaction. In the present case, $a_{{\rm IB}}$ and $a_{{\rm BB}}$ present the
same dependence on the polarization angle, according to
Eq.~(\ref{scatt_length}).  Consequently, for fixed impurity and bath,
the ratio $a_{{\rm IB}}/a_{\rm BB}=\beta$ remains constant when $\alpha$
changes, and the product $E_p^{(0)}(\alpha) a^2_{\rm BB}(\alpha)$ becomes
a function of the gas parameter $x = n a_{\rm BB}^2$ alone. This means that changing the polarization angle $\alpha$ leaves the mean-field polaron energy unchanged if the density is changed accordingly to keep the gas parameter constant.

Considering this property emerges from Eq.~(\ref{Emf_polaron}), it
is in principle expected to hold only at low $x$. However, and surprisingly, it is still present up to very large values of the gas parameter $x \sim 100$, as shown in Fig.~\ref{fig:fig2}. We show in the figure the DMC results for the polaron energy, expressed in scattering length units, i.e. $E_p(\alpha)/(\hbar^2/ma_{\rm BB}^2(\alpha))$, as a function of the tilting angle for three different values of the gas parameter ($x=0.001$, $1$, $10$) and a fixed ratio $\beta=a_{{\rm IB}}/a_{\rm BB}=10$. These parameters place our system away from the perturbative regime, meaning that the background gas of Bosons is highly correlated and the impurity-bath interaction can not be considered a perturbation.
From the results, we see that the quantity $E_p(\alpha)/(\hbar^2/ma_{\rm BB}^2(\alpha))$ is kept constant (except for small variations of less than $5 \%$) when $\alpha$ changes for a fixed gas parameter.
Thus, in a very good approximation, given a fixed value of $\beta$, the polaron energy is a function of the density $n$ and the boson-boson $a_{\rm BB}$ (or impurity-boson $a_{\rm IB}$) scattering length alone for different tilting angles, meaning that different dipolar interactions with different tilting angles display universality. The deviations from a constant of the rescaled polaron energy $E_p(\alpha)/(\hbar^2/ma_{\rm BB}^2(\alpha))$ grow with $\alpha$ for the highest tilting angles considered, which indicates that universality starts to slightly break down in the limit $\alpha \rightarrow 0.615$ rad, whence the DDI acquires a negative contribution that induces a collapse into the system. This can be understood as follows: increasing $\alpha$ while keeping the gas parameter constant implies progressively reducing the s-wave scattering length while increasing the density, which in turn decreases the inter-particle distance and thus enhances finite range effects.
These results align with the previous findings for the bulk
system, where a similar universal behavior is
observed~\cite{SanchezBaena2022}. In this sense, the universality present in this work is very much akin to the one discussed in Ref.~\cite{durst24}, where it is shown that the absorption spectra of a Rydberg atom is a function solely of the parameter $n a^3$, despite the Rydberg potential depending on the principal quantum number, which controls its depth and its range. This means that all Rydberg interactions show the same universal response which only depends on $n$ and $a$, in much the same way that different dipole-dipole interactions with different tilting angles display results that only depend on $n$ and $a_{\rm BB}$, given that $\beta$ is kept fixed.

Given this trend displayed by the polaron energy and the
analytic expression of the scattering length in
Eq.~(\ref{scatt_length}), the dipolar polaron energy can be considered
to be a function of $\beta$, $n$ and $a_{\rm BB}$ alone that can be obtained from a fit to the
corresponding Monte Carlo data. We have obtained these fits for
$\alpha=0$ and
two relevant values of the coupling strength: $\beta=1.42$,
corresponding to the case of a Dy impurity immersed in an Er bath, and
the extreme case of a strongly interacting impurity, $\beta=10$. We
have checked that the polaron energy follows a law of the form
\begin{equation}
 E_p(\alpha=0) = \exp( a(\ln(x) + c)^d + b ) \epsilon_d \ ,
 \label{eqn:fit}
\end{equation}
with $a = 0.94(8)$, $b = -16.30(1)$ for $\beta=1.42$ and $a = 0.97(4)$,
$b =-15.79(4)$ for $\beta=10$. In both cases, $c = 11.99(3)$, and
$d = 1.09(3)$. Also, $x = n a_{\rm BB}^2$ and $\epsilon_d = \hbar^2/m d_B^2$, where $d_{\rm B} = m_{\rm B}
C_{dd}^{\rm BB}/ (12 \pi \hbar^2)$ is the dipolar length of the atoms of the bath. The energy for any other tilting angle can approximately be obtained by the relation
\begin{align}
 E_p(\alpha) \simeq E_p(\alpha=0) \frac{ \left( \hbar^2/(m a_{\rm BB}^2(\alpha)) \right) }{ \left( \hbar^2/(m a_{\rm BB}^2(\alpha=0)) \right) } \ .
 \label{eqn:univ_energy}
\end{align}
The functional form of Eq.~\ref{eqn:fit} is phenomenological, in the sense that it has not been derived from any first principles, but rather has been chosen such that the DMC results for $\alpha=0$ can be reproduced with a small error. In all cases this error is slightly less than $10\%$.


A relevant question related to the previous results is the extent to
which a perturbative approximation accurately describes the ground
state energy of the system for the dipolar polaron.
In a perturbative scheme, 
the bath is usually described by a Bogoliubov Hamiltonian in the
absence of the impurity, while the impurity-bath interaction is
considered to be the (weak) perturbation.
For the two-dimensional dipolar system considered in this
work, 
the boson-boson and impurity-boson interactions in momentum space are
taken to be described by the pseudopotentials~\cite{macia12}
\begin{align}
 V^{(p)}_{\sigma\sigma'}({\bf k}) &= -\frac{4 \pi \hbar^2}{m
   \ln(n a_{\sigma\sigma'}^2) }
 + \frac{C_{dd}^{\sigma\sigma'} k \pi \sin^2 \alpha \cos 2 \theta_k}{2}
 \label{pseudopotential}
\end{align}
with $\theta_k$ the polar angle of the momentum vector. This
pseudopotential is built such that its s-wave and d-wave scattering properties computed under the first order Born approximation match those of the full dipole-dipole interaction, i.e. Eq.~\ref{eq:V12}.
Note also that this pseudopotential incorporates finite-range effects
via the anisotropic contribution in the second term. Within
perturbation theory and using the Fröhlich Hamiltonian, one
considers only
processes
where the impurity couples to a single
excitation of the medium at once.  In order to quantify the accuracy
of the perturbative approach, we show in Fig.~\ref{fig:fig3}(a) the
ratio of the DMC energies
to the lowest order perturbation prediction $E_p^{(0)}$ of
Eq.~(\ref{Emf_polaron}).
As it can be seen from the figure and as expected, the
perturbative approximation holds only in the dilute limit, corresponding
to gas parameter values $x \lesssim 0.01$ for $\beta = 1.42$
and $x \lesssim 0.001$ for $\beta = 10$. For larger
values of $x$, higher-order effects, neglected in the
lowest-order perturbative scheme, start to become important.

Finally, Fig.~\ref{fig:fig3}(b) shows the dependence of the polaron
energy on the coupling ratio $\beta$ for $\alpha=0$ and several values
of the gas parameter. Energies have been rescaled with respect to
their values at $\beta=1$ for the sake of comparison, which
correspond to $E_p(x=10^{-5}) = 1.25 \times 10^{-7}\epsilon_d$,
$E_p(x=10^{-2}) = 3.92 \times 10^{-4}\epsilon_d$ and $E_p(x=10^{2}) =
58.18 \epsilon_d$.  We find that the relative variation of the polaron
energy grows with increasing gas parameter. This is a consequence of
the fully repulsive character of the dipole-dipole interaction and the
fact that, for a fixed polarization angle, increasing the value of the gas
parameter is equivalent to increasing the density of atoms of the
bath. We show only the variation of the polaron energy with respect to $\beta$ at zero tilting angle because the energy for any other value of $\alpha$ can be obtained through Eq.~\ref{eqn:univ_energy} due to the universality of the results with respect to the tilting angle.

\begin{figure*}
\includegraphics[width=0.6\textwidth]{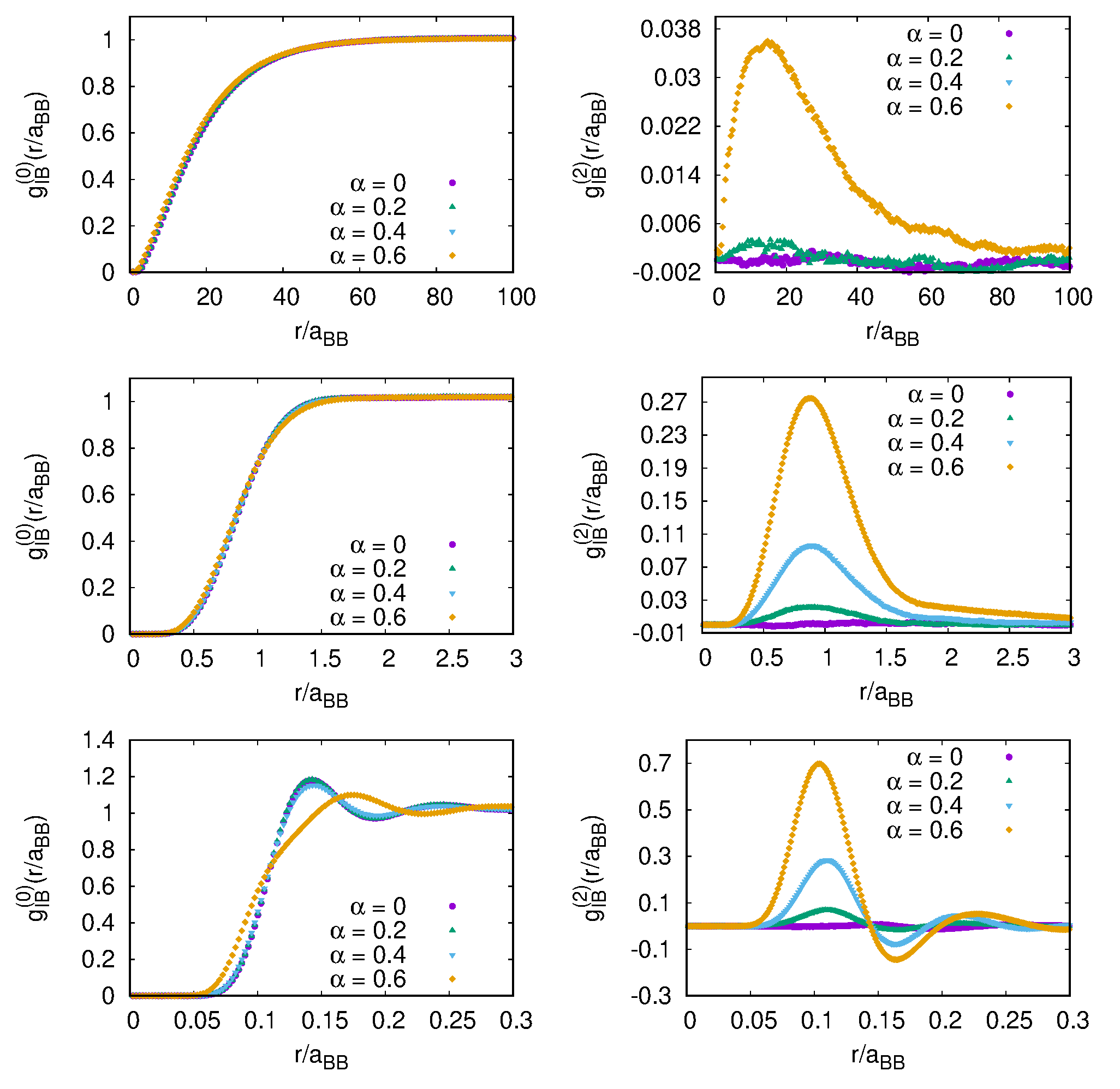}
\caption{
  $s$- and $d$- partial wave modes of the
  impurity-boson pair-distribution function (dimensionless) for three characteristic
  values of the gas parameter: $x=0.001$ (top row), $x=1$ (middle
  row), $x=100$ (bottom row).  The impurity strength ratio is fixed to
  $\beta=10$. The tilting $\alpha$ is expressed in radians.
}
\label{gr_impurity}
\end{figure*}

Additional insight into the dipolar polaron universality can be
drawn from the relation between the polaron energy and the boson-boson
and impurity-boson pair distribution functions. These quantities are
defined as
\small
\begin{eqnarray}
  g_{\mathrm{BB}}({\bf r_1}-{\bf r_2})
  & = & 
  \frac{N(N-1)}{n^2}
  \frac{\int d{\bf r_3} \cdots d{\bf r_N} d{\bf r_I}
    \abs{\Psi({\bf R})}^2 }{ \int d{\bf R} \abs{\Psi({\bf R})}^2 }
  \\
  g_{\mathrm{IB}}({\bf r_I}-{\bf r_1}) & = &
  \frac{N}{n_I n} \frac{\int d{\bf r_2} \cdots d{\bf r_N}
    \abs{\Psi({\bf R})}^2 }{ \int d{\bf R} \abs{\Psi({\bf R})}^2 } \ ,  
\end{eqnarray}
\normalsize
with ${\bf R}$ representing the set of all particle coordinates,
$n=N/V$ and $n_I = 1/V$ being the average bath and impurity density,
respectively.  Actually, these two functions can be expanded in
partial waves and, due to the anisotropy of the dipolar interaction,
they present non-zero contributions beyond the s-wave.  We show in
Figure~\ref{gr_impurity} the first two
modes of
\begin{equation}
  g_{\mathrm{IB}}({\bf r_I}-{\bf r_1}) = \sum_{l=0}^{\infty}
  g^{(2l)}_{\mathrm{IB}}(r) \cos 2 l \theta
\end{equation}
for $\beta=10$ and different gas parameters and tilting angles.
Results for $g_{\mathrm{BB}}({\bf r_1}-{\bf r_2})$ are very similar to
those obtained for the dipolar bulk case in
Ref.~\cite{SanchezBaena2022}.  As it can be seen from the figure,
the isotropic mode is universal with respect to the tilting angle up to $\alpha \simeq 0.4$, since all curves collapse to a single one when plotted in terms of the rescaled distance $r/a_{\rm BB}$. This happens even for
abnormally large values of the gas parameter. On the other hand, the first
anisotropic mode does not show any universality in $\alpha$ at all. However, we
also see that, up to $\alpha \simeq 0.4$, the isotropic mode clearly
dominates over the anisotropic contribution unless both $\alpha$ and
$x$ are large, leading to an essentially universal behaviour,
similarly to what
it was reported
in~\cite{SanchezBaena2022} for the bulk.  The pair distribution
functions is related to the potential energy of the system by the
relation
\begin{figure}[t]
\includegraphics[width=0.5\textwidth]{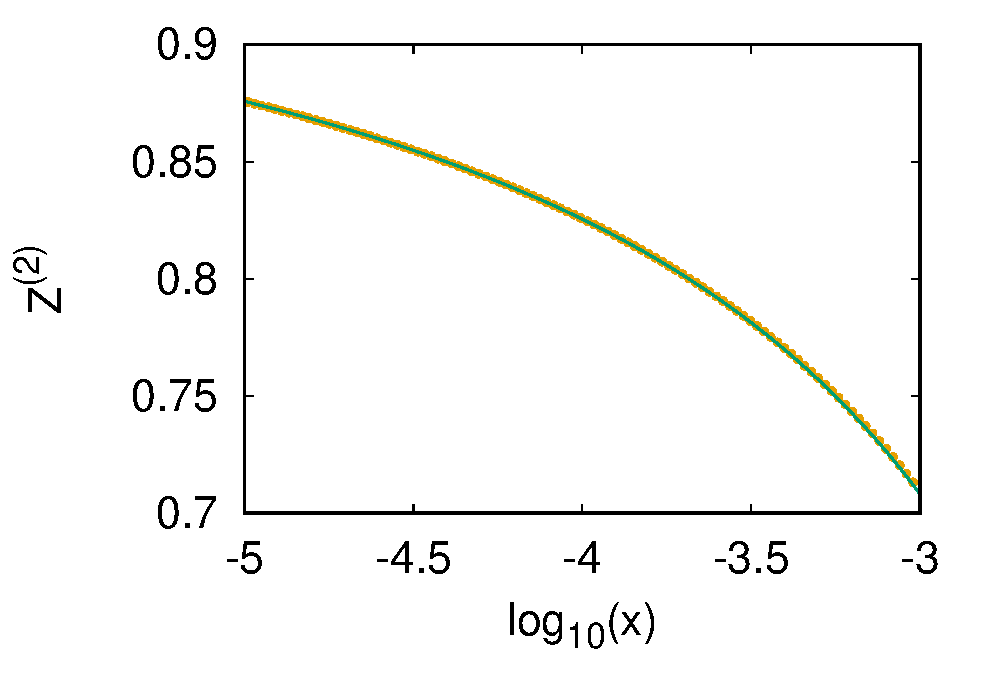}
\caption{ Second order perturbation theory results for the quasiparticle residue $Z$ (dimensionless) for a bath with (orange dots) and without (green solid line) the finite range, anisotropic contribution of the boson-boson interactions (see Eq.~\ref{pseudopotential}). In both cases, $\alpha=0.6$ rad for the impurity-boson interaction.}
\label{Z_polaron_new}
\end{figure}

\begin{align}
  \langle V \rangle = n \int d{\bf r}V_{\mathrm{IB}}({\bf r}) g_{\mathrm{IB}}({\bf r}) +
  \frac{nN}{2} \int d{\bf r}V_{\mathrm{BB}}({\bf r}) g_{\mathrm{BB}}({\bf r}) \ ,
\end{align}
where the first term comes from the interaction between the impurity
and the rest of the particles in the medium, while the second term
accounts for the contribution of the bath. From this expression, one
can recover the total energy of the system through the
Hellmann-Feynman theorem~\cite{SanchezBaena2022,Fetter}
\begin{align}
  E & = \int_0^1 du \left\{ n \int d{\bf r}V_{\mathrm{IB}}({\bf r})
        g_{\mathrm{IB}}({\bf r},u) \right.
  \nonumber \\
        & \left. + \frac{nN}{2} \int d{\bf r}V_{\mathrm{BB}}({\bf r})
        g_{\mathrm{BB}}({\bf r},u) \right\}
\end{align}
where $g_{\mathrm{IB}}({\bf r},u)$ and $g_{\mathrm{BB}}({\bf r},u)$
stand the pair distribution functions corresponding to the Hamiltonian
$\hat{H} = \hat{H}_{kin} + \hat{H}_{pot} u$, with $\hat{H}_{kin}$ and
$\hat{H}_{pot}$ the kinetic and potential terms of the Hamiltonian in
Eq.~(\ref{eq:HFull}), respectively. The polaron energy can then be
recovered from 
the energy difference
\begin{align}
  & E_p = E(N, 1) - E(N, 0) = \int_0^1 du
    \left\{ n \int d{\bf r}V_{\mathrm{IB}}({\bf r})
    g_{\mathrm{IB}}({\bf r},u) \right.
  \nonumber \\
  & \left. + \frac{nN}{2} \int d{\bf r}V_{\mathrm{BB}}({\bf r})
    g_{\mathrm{BB}}({\bf r},u) \right\} - \frac{nN}{2}
    \int d{\bf r}V_{\mathrm{BB}}({\bf r}) \tilde{g}_{\mathrm{BB}}({\bf r},u)
\end{align}
where $E(N_B,N_I)$ denotes the ground-state energy of a system with
$N_B$ bosons and $N_I$ impurities, and $\tilde{g}_{\mathrm{BB}}({\bf
  r},u)$ is the boson-boson pair distribution function of the bulk
system (i.e. for absent impurity).  In this way, the universality in
the polaron energy can be understood as being inherited from the
corresponding behavior of the pair distribution functions.

\begin{figure*}[t]
\includegraphics[width=1\textwidth]{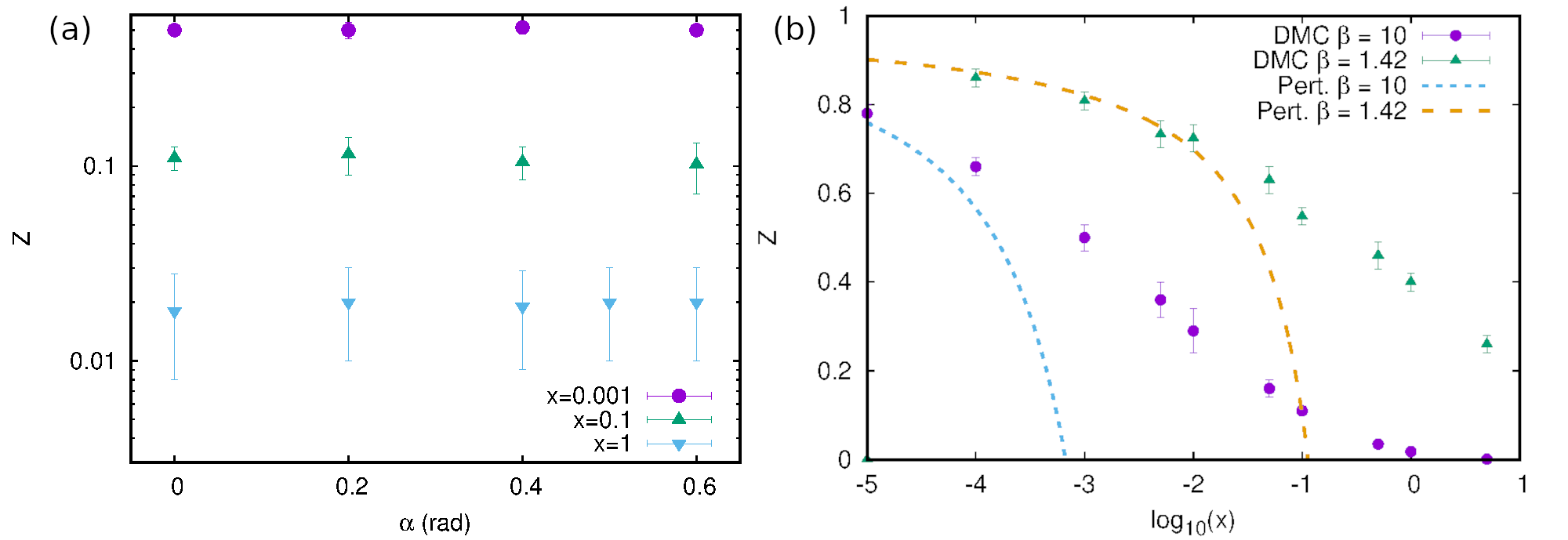}
\caption{ (a) DMC results for the quasiparticle residue (dimensionless) as a function of the tilting angle for several values of the gas parameter for $\beta=10$. (b) DMC results (dots) and perturbative results (solid lines) for $Z$ obtained for $\alpha=0$ rad as a function of the gas parameter.}
\label{Z_polaron}
\end{figure*}



\subsection{Quasiparticle residue}

Another experimentally relevant quantity in the study of the polaron
physics is the quasiparticle residue $Z$, which
quantifies the overlap between the full wave function of the system
and a state conformed by a non-interacting impurity and a vacuum of
excitations. A mixed estimator for this quantity can be obtained in
DMC from the long-range asymptotic behaviour of the one-body density
matrix associated to the impurity~\cite{ArdilaPRR2020,maciathesis}
\begin{equation} Z=\lim_{r\rightarrow\infty}\rho(\mathbf{r})=\lim_{r\rightarrow\infty}\left\langle \frac{\psi_{T}(\mathbf{r}_{\mathrm{I}}+\mathbf{r},\mathbf{r}_{1},\cdots,\mathbf{r}_{N})}{\psi_{T}(\mathbf{r}_{\mathrm{I}},\mathbf{r}_{1},\cdots,\mathbf{r}_{N})}\right\rangle
\label{pert_Z}
\end{equation}
where $\psi_T$ is the many-body trial wave function guiding the
simulation. We report in Fig.~\ref{Z_polaron}(a) the dependence of $Z$
on the polarization angle $\alpha$ for different values of the gas
parameter and $\beta=10$.
As can be seen, $Z$ is also independent of $\alpha$ and 
seems to depend on the gas parameter exclusively, even for the
largest values of $x$ where inter-atomic correlations play an
important role. In other words, the quasiparticle residue shows a clear universal behaviour with respect to the tilting angle. This surprising property can be hinted already at the
perturbative level
using the simple model described above.
To second order and using the interaction in
Eq.~(\ref{pseudopotential}) one finds
\begin{align}
  Z^{(2)} = \left( 1 + \frac{n}{(2 \pi)^2}
  \int d{\bf k} (V^{(p)}_{\mathrm{IB}}({\bf k}))^2
  \frac{\epsilon_{{\bf k}}}{E({\bf k})}
  \frac{1}{ \left( \epsilon_{{\bf k}} - E({\bf k}) \right)^2} \right)^{-1}
\end{align}
with $\epsilon_{{\bf k}} = \frac{\hbar^2 k^2}{2m}$ and $E({\bf k}) =
\sqrt{ \epsilon_{{\bf k}} \left( \epsilon_{{\bf k}} + 2 n
  V^{(p)}_{\mathrm{BB}}({\bf k}) \right) }$ the excitation spectrum of
the bulk. Interestingly, getting rid of the anisotropic contribution to
the bath-bath pseudopotential (second term on the rhs of
Eq.~(\ref{pseudopotential})) that enters $Z$ through the excitation
spectrum of the medium leaves the quasiparticle residue essentially
unchanged. This is shown in Figure~\ref{Z_polaron_new}, where we
compare the values of $Z$ obtained with and without this contribution
for different values of the gas parameter and $\alpha=0.6$.
In this way, 
the only
relevant anisotropic contribution to $Z$
comes from the impurity-bath
interaction.
However, the lowest order anisotropic contribution is proportional to 
$\sin^4 \alpha \ll 1$, since the term proportional to $\sin^2 \alpha
\cos 2\theta_k$ coming from $(V^{(p)}_{\mathrm{IB}}({\bf k}))^2$
yields zero contribution when the angular integration is
evaluated. This means that, at the perturbative level,
the dependence of
$Z$ on the density and the impurity-bath scattering length is the same
as that of an isotropic system with zero range interactions,
and thus $Z$ is a function of the gas parameter alone~\cite{ArdilaPRR2020}.

At this point, one can also compare the DMC prediction of the
quasiparticle residue to the results obtained with perturbation
theory, as a way to benchmark the perturbative approach. We show in Fig.~\ref{Z_polaron} (b) the DMC estimation of $Z$
together with the perturbative result obtained from Eq.~(\ref{pert_Z})
for the two cases $\beta=1.42$ and $\beta=10$ and $\alpha=0$.  Because
of universality with respect to the polarization angle, the same results hold for any other value of
$\alpha$. As it can be seen, the predictive power of the perturbative
approach worsens with increasing $x$ and/or $\beta$, as happens with
the polaron energy.  In this case, though, the situation is worse as
the perturbative prediction ceases to reproduce the DMC data at lower
gas parameter values, at least for $\beta=10$.
Remarkably, for the lowest coupling
case, the regime where $Z$ is close to unity extends up to $x \lesssim 10^{-2}$.

\begin{figure}[b]
\includegraphics[width=0.5\textwidth]{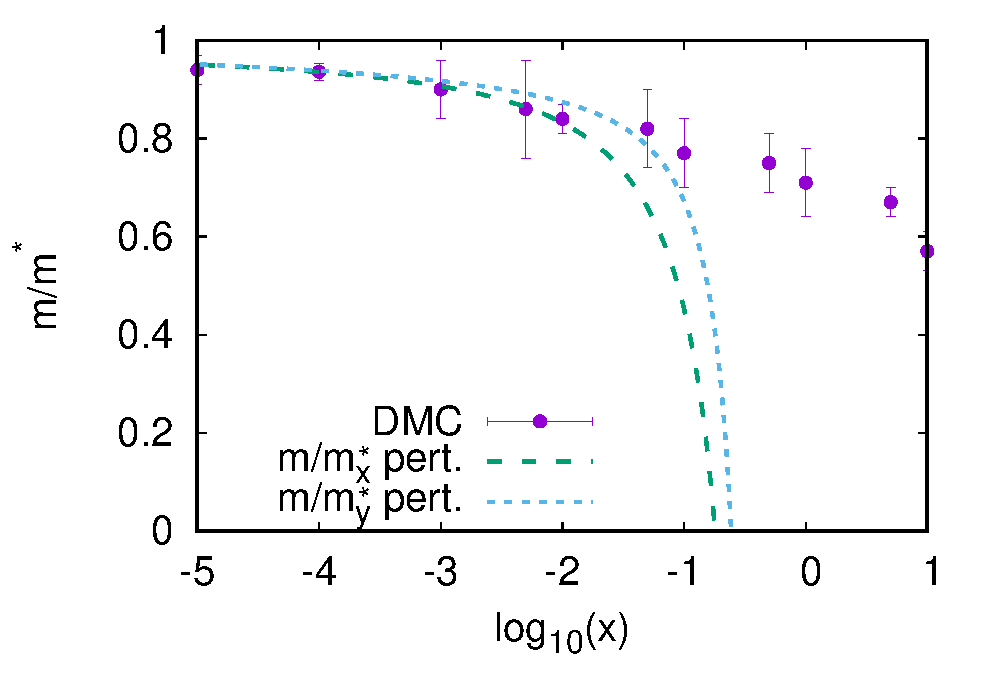}
\caption{DMC (dots) and second order perturbation theory (solid line) results for the inverse effective mass (dimensionless) as a function of the gas parameter for $\alpha=0.6$ rad, $\beta=1.42$. DMC results correspond to $m/m_x^* \simeq m/m_y^*$ since, in this regime, the DMC estimations of $m_x^*$ and $m_y^*$ are indistinguishable within statistical noise.
%
}
\label{m_m_eff_aniso}
\end{figure}

\subsection{Effective mass}

The last quantity we address in this work is the polaron's effective
mass.  In order to obtain the effective mass in DMC, one can track the
diffusion movement of the polaron in imaginary time. This is done
calculating its mean-square displacement according to the expression
$\frac{m}{m^\ast}=\lim_{\tau\rightarrow\infty}\frac{\left\langle
  \left|\Delta\mathbf{r}_{\mathrm{I}}(\tau)\right|^{2}\right\rangle
}{4D\tau}$~\cite{ArdilaPRR2020,bombin19}; with $D=\hbar^2/(2m)$ the
diffusion constant of a free particle, and $\langle \left|\Delta{\bf
  r}_{\mathrm{I}}(\tau)\right|^{2}\rangle =\langle \left|{\bf
  r}_{\mathrm{I}}(\tau)-{\bf r}_{\mathrm{I}}(0)\right|^{2}\rangle$,
$\tau=it/\hbar$ being the imaginary time of the simulation.
Due to the anisotropic character of the dipolar interaction,
the effective mass turns out to depend on the impurity's momentum
direction for $\alpha\neq 0$. In order to quantify this effect, one
can define an anisotropic effective mass by tracking the position of
the impurity in each direction separately, according to
$\frac{m}{m_\chi^\ast}=\lim_{\tau\rightarrow\infty}\frac{\left\langle
  \left|\Delta\chi_{\mathrm{I}}(\tau)\right|^{2}\right\rangle
}{2D\tau}$ with $\chi = x \text{ or } y$.
Notice that the previous expression contain a factor of $2$ instead
of $4$ in the denominator,
as in this case the diffusion is treated as
one-dimensional.

\begin{figure*}[t]
\includegraphics[width=1\textwidth]{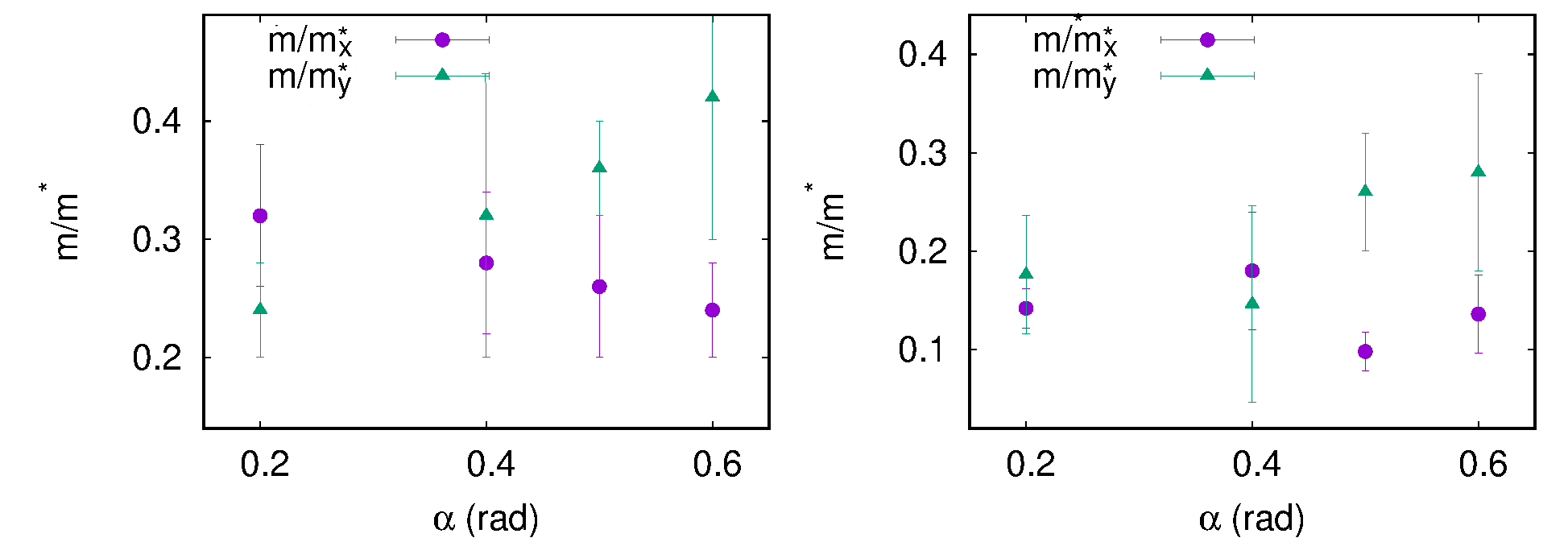}
\caption{DMC Results for the inverse effective mass (dimensionless) for $x=1$ (a) and $x=100$
  (b) as a function of the tilting angle. In both cases, $\beta=10$.}
\label{m_m_eff_aniso_dmc}
\end{figure*}

In any case and as done before, it is interesting to discuss first the
predictions of perturbation theory.
To second order,
the effective mass is obtained from the
second-order
polaron energy~\cite{Ardila:2018jo}
\begin{align}
  E^{(2)}_p = \frac{ \bra{ {\bf P}, 0 }
    \hat{H}_{\mathrm{IB}} \ket{ \Psi_1 } }{ \bra{\Psi} \ket{\Psi} }
 \label{secondorderE}
\end{align}
where $\bra{ {\bf P}, 0 }$ denotes a state with a non-interacting
impurity with momentum ${\bf P}$ and a zero-momentum medium. In much
the same way, $\ket{\Psi} = \ket{ {\bf P}, 0 } + \lambda \ket{ \Psi_1
}$ with $\lambda$ a perturbative parameter proportional to the
strength of the impurity-medium interaction and $\ket{ \Psi_1}$ the first order contribution to the total wave function accounting for the perturbation.

Evaluating Eq.~(\ref{secondorderE}) and performing a Taylor
expansion in terms of the impurity momentum, the $P^2$ contribution
results in the following momentum-dependent correction to the polaron energy
\small
\begin{align}
 \Delta E = -\frac{P^2}{2 m}\frac{1}{(2\pi)^2} \int d{\bf k} \text{ }
 n \left( V^{(p)}_{\mathrm{IB}}({\bf k}) \right)^2 \frac{\epsilon_{\bf
     k}}{E_{\bf k}} \frac{ 2 \hbar^2 k^2 }{\left( \epsilon_{\bf k} +
   E_{\bf k} \right)^3 } \cos^2 \theta
 \label{secondorderE_2}
\end{align}
\normalsize
where $\theta$ is the angle between the impurity momentum ${\bf P}$
and the integration momentum $\hbar{\bf k}$, which forms an angle $\phi$ with the $x$ axis.
The effective mass is then
obtained from the momentum-dependent part of the polaron energy, which
is given by the sum of impurity kinetic energy and the correction in
Eq.~(\ref{secondorderE_2}), i.e.
\begin{align}
\frac{P^2}{2 m^*} &= \frac{P^2}{2 m} + \Delta E = \frac{P^2}{2 m} \left( 1 + \frac{2 m \Delta E}{P^2} \right) \nonumber \\
&= \frac{P^2}{2 m} \left( 1 - \frac{1}{(2\pi)^2}
\int d{\bf k} \text{ } n \left( V^{(p)}_{\mathrm{IB}}({\bf k})
\right)^2 \right. \nonumber \\
&\left. \times \frac{\epsilon_{\bf
    k}}{E_{\bf k}} \frac{ 2 \hbar^2 k^2 }{\left( \epsilon_{\bf k} +
  E_{\bf k} \right)^3 } \cos^2 \theta \right) \ .
 \label{effective_mass_pert}
\end{align}
This result indicates that the inclusion of the second term in the
pseudopotential of Eq.~(\ref{pseudopotential}) leads to an anisotropic
effective mass, induced by the angular dependence on the impurity's
momentum vector. In particular, for an impurity moving along the $x$ axis, $m^*$ is replaced by $m^*_x$ in Eq.~(\ref{effective_mass_pert}) and this results into the substitution $\cos^2 \theta \rightarrow \cos^2 \phi$ while for a impurity moving along the $y$ axis, computing $m^*_y$ implies setting $\cos^2 \theta \rightarrow \sin^2 \phi$.
Furthermore and as happens with the quasiparticle residue,
the anisotropy of the boson-boson interactions that enters
Eq.~(\ref{effective_mass_pert}) through the excitation spectrum of the
background has little
impact on $m^*$ compared to the impurity-boson potential.
Since, as usual, the
second order
correction to the polaron energy is negative,
$m^*>m$
and the impurity acts as a heavier quasiparticle in the medium.

Figure~\ref{m_m_eff_aniso} shows the ratios $m/m_x^{*}$ and $m/m_y^{*}$ obtained with
second order perturbation theory for $\beta=1.42$ and $\alpha=0.6$ as
a function of the gas parameter. As one can see, the effective mass is
larger when the polaron moves along the $x$ axis. This is because the
anisotropy in the effective mass is determined by the anisotropy of
the impurity-bath interaction in momentum space, which is maximally
repulsive along this direction.
The corresponding DMC predictions for the effective mass in the $x$ (or $y$) axis are also shown in the plot.
Because the noise in the estimation in the effective mass is large,
it prevents a clear observation of its anisotropic character but at
large enough gas parameters and impurity-bath coupling strengths. This implies that, in this regime, the DMC results for the effective mass along the $x$ and $y$ axes are indistinguishable up to statistical noise, which is why only a set of points is shown.
This is an issue even when long simulations, which accumulate a large
quantity of statistical data, are performed. Regardless, we see
agreement between the perturbative and DMC results in the regime where
$m_x^* \simeq m_y^*$.

The anisotropic character of the effective mass is more clearly seen
when correlations are strong. In order to showcase that, 
we show in Fig.~\ref{m_m_eff_aniso_dmc} the DMC results for
$m/m_x^{*}$ and $m/m_y^{*}$ as a function of the polarization angle
for $\beta=10$ and two values of the gas parameter,
$x=1$ and $x=100$.
We can see that, even away from the regime of validity of perturbation
theory ($\beta \gg 1$), anisotropic effects in the effective mass
follow the qualitative trends predicted by the perturbative
calculation, showing indeed that $m_x^*>m_y^*$.

\section{Experimental viability}

Even though some of the gas parameters considered in this work are too high to be experimentally viable, the universality with respect to the tilting angle in the polaronic properties takes place also at moderately high gas parameters $x \sim 0.01$, which lay within the reach of potential experiments and are above, for instance, of the threshold for which universality is lost for the polaron energy of the two-dimensional Fermi polaron, $x \sim 10^{-5}$~\cite{bombin19}. As an example, let us consider a system of Dy atoms of 2D density $n_{2D}$ in the $x$-$y$ plane. In a realistic experiment, atoms which lie within a 3D set-up lay in the 2D regime if $L_z < a_{3D}$~\cite{pilati05}, with $L_z$ the size of the system along the $z$-axis and $a_{3D}$ the 3D scattering length. In experiments, $a_{3D}$ is of the order of the dipolar length $a_{\rm dd}$, defined as $a_{\rm dd} = m C_{\rm dd}/(12 \pi \hbar^2)$. Consider a 2D gas parameter of $x=0.01$ with a 2D scattering length of $a_{2D} =962.6 a_0$ (which corresponds to $\alpha=0.4$ for $^{164}$Dy atoms) with $a_0$ the Bohr radius. Taking $L_z = 0.5 a_{\rm dd}$, which guarantees that the 3D system is in the 2D regime, since $a_{3D}$ should be of the order of $a_{\rm dd}$ or larger, yields a 3D density of $n_{3D} = 1.1 \times 10^{15}$ cm$^{-3}$. This density is not far from those present in experiments of dipolar gases~\cite{FerrierBarbut:2016jo}, meaning that this regime where universality is not expected but is reported in our study could be experimentally probed. Moreover, for this particular case, the presence of three-body losses would not be an issue for the formation of the polaron and the subsequent probing of polaronic properties. According to Ref.~\cite{smith23}, the characteristic three-body lifetime can be estimated as $t_{3} \sim 1/(L_3 n^2)$, with $L_3$ the three-body loss coefficient, which for Dy atoms is measured to be $L_3 = 1.33 \times 10^{-41} $ m$^6$$/$s~\cite{bottcher19}. For the 3D density considered, this yields $t_3 \simeq 69.3$ ms. This lifetime is significantly higher than the time it takes for the polaron to be formed, which is of the order of $\mu$s in experiments for non-dipolar atoms~\cite{skou2021} and of the order of $0.1$ ms for dipolar ones~\cite{volosniev23}.

\section{Conclusions and future perspectives}
To summarize, we have studied the quasiparticle properties of a
dipolar impurity immersed in a dipolar bath in two dimensions, both
being subject to an external polarization field that makes all
dipole moments point in the same direction.
In order to do that, we have used the Diffusion Monte Carlo (DMC)
method, comparing the results to second order perturbation theory.
We have shown that, to a large extent, and for a fixed ratio of the boson-boson and impurity-boson scattering lengths, the polaron energy displays universal behaviour with respect to the tilting angle, since it is only a function of the density $n$ and the boson-boson $a_{\rm BB}$ (or impurity-boson $a_{\rm IB}$) scattering length.
This is directly induced by the same universal behaviour of the pair-distribution
function, where the isotropic mode dominates. We have also shown that
the quasiparticle residue shows also universality with respect to the tilting angle, a result that is recovered through perturbation theory even
when anisotropic finite range effects are considered. Finally, we have
shown that the anisotropy of the dipole-dipole interaction leads to an
anisotropic effective mass which,
surprisingly,
is larger in the direction of minimum repulsion of the dipole-dipole
interaction in position space, which is a consequence of its angular
dependence in momentum space. For all the aforementioned properties,
we have established the regime of validity of perturbation theory in
terms of the gas parameter by direct comparison to the DMC results.

Interesting directions for future research include the exploration of the attractive polaron branch and the formation of many-body Efimov states. This is not possible with the inter-particle potential employed in the simulations of this work, since it lacks any attractive component, but could be done by modelling the inter-particle interaction by a dipole-dipole potential plus a short range repulsive core, and allowing for tilting angles $\alpha > 0.615$ rad. On top of that, the interplay between supersolidity and the polaron could be studied following this same route, since this inter-particle interaction leads to the formation of a striped dilute liquid~\cite{staudinger23}.

%

\begin{acknowledgements}
The work has been supported by grant PID2020-113565GB-C21 from
MCIN/AEI/10.13039/ 501100011033. L.A.P.A acknowledges the support of the
Deutsche Forschungsgemeinschaft (DFG, German Research Foundation)
under Germany’s Excellence Strategy– EXC-2123 QuantumFrontiers–
390837967, and FOR 2247.  L.A.P.A also acknowledges by the PNRR MUR
project PE0000023 - NQSTI and the Deutsche Forschungsgemeinschaft
(DFG, German Research Foundation) under Germany’s Excellence Strategy
- EXC - 2123 Quantum Frontiers-390837967 and FOR2247
\end{acknowledgements}

\bibliography{references}

\bibliographystyle{apsrev4-1}

\clearpage

\widetext

\renewcommand{\theequation}{S\arabic{equation}}
\renewcommand{\thefigure}{S\arabic{figure}}
\renewcommand{\thetable}{S\arabic{table}}

\onecolumngrid

\end{document}